\documentclass{PoS}

\title{ElectroWeak BaryoGenesis via Top Transport}

\ShortTitle{EWBG2-tt/tc}

\author{\speaker{George W.-S. Hou}
   \thanks{Work done in collaboration with Kaori Fuyuto and Eibun Senaha.}\\
        University of Melbourne,
         and National Taiwan University\footnote{
             Home institute.
             }
        \\
        E-mail: \email{wshou@phys.ntu.edu.tw}}

\abstract{
We study electroweak baryogenesis driven by the top quark in 
two Higgs doublet model that allows flavor-changing neutral Higgs couplings.
Taking Higgs sector couplings and the additional top Yukawa coupling $\rho_{tt}$
to be $\mathcal{O}$(1), one naturally has first order electroweak phase transition
and sufficient $CP$ violation to fuel the cosmic baryon asymmetry.
Even if $\rho_{tt}$ vanishes, the favor-changing coupling $\rho_{tc}$
can still achieve baryogenesis.
Phenomenological consequences such as $t\to ch$, $\tau \to \mu\gamma$,
electron electric dipole moment, $h\to\gamma\gamma$,
and $hhh$ coupling are discussed.
The extra scalars $H^0$, $A^0$ and $H^\pm$ are sub-TeV in mass,
and can be searched for at the LHC.
}

\FullConference{EPS-HEP 2017, European Physical Society conference on High Energy Physics\\
		5-12 July 2017\\
		Venice, Italy}

\begin{document}

\section{Introduction: ``General'' 2HDM}

Remarkably, the Standard Model (SM) carries all ingredients of
the Sakharov conditions for generating baryon asymmetry of the Universe (BAU),
albeit falling short on order of electroweak phase transition (EWPT),
and strength of $CP$ violation (CPV):
the weak interaction is too weak to make EWPT 1st order, 
while the Jarlskog invariant is too meager a source for CPV.
Involving all three generations, the latter is 
suppressed by both small masses and mixing angles.

If one adds a second Higgs doublet (2HDM), it is known that 
one could have 1st order EWPT if Higgs sector couplings are ${\cal O}(1)$.
For CPV, while some have tried complex couplings in Higgs sector,
it may be prudent to keep the Higgs potential
$CP$ invariant to avoid trouble with neutron edm, $d_n$.
Recalling that known CPV arises from the CKM matrix, i.e. from Yukawa couplings,
one naturally asks whether there can be extra Yukawa couplings in 2HDMs.
Alas, such couplings were eliminated by the Natural Flavor Conservation 
(NFC) condition of Glashow and Weinberg~\cite{Glashow:1976nt} 40 years ago.
NFC is usually implemented by imposing a $Z_2$ symmetry 
on the two Higgs fields $\Phi_1$ and $\Phi_2$
to forbid flavor-changing neutral Higgs (FCNH) couplings.
Admittedly, such discrete symmetries may seem \emph{ad hoc},
and indeed deemed perhaps unnecessary~\cite{Cheng:1987rs}, given the observed
trickle-down pattern or mass suppression of far off-diagonal quark mixings.

Here, we drop $Z_2$ symmetry (or NFC) and utilize extra Yukawa couplings 
$\rho_{tt}$ and $\rho_{tc}$, which are naturally ${\cal O}(1)$ and complex, 
to drive~\cite{Fuyuto:2017ewj} EWBG.
We note that, recently, many authors have taken a data-driven
approach to these FCNH couplings,
not just in the old suggestion~\cite{Hou:1991un} of $t\to ch$ decay, but applying 
also to the so-called $B \to D^{(*)}$ anomaly, as well as $h \to \tau\mu$ decay.

\section{Model}

The Yukawa interaction for up-type quarks
in a general 2HDM without $Z_2$  is
\begin{eqnarray}
-\mathcal{L}_Y
 = \bar{q}_{iL}\left(Y^u_{1ij}\tilde{\Phi}_1+Y^u_{2ij}\tilde{\Phi}_2 \right) u_{jR}+{\rm h.c.},
\end{eqnarray}
where $i, j$ are flavor indices and $\tilde \Phi_b = i\tau_2\Phi^*_b$ ($b=1, 2$).
With $\Phi_{1,2}$ each acquiring a vacuum expectation value (VEV) $\upsilon_{1,2}$,
and defining the usual $\upsilon_1 = \upsilon c_\beta$, $\upsilon_2 =\upsilon s_\beta$
(hence $\upsilon^2 = \upsilon_1^2 + \upsilon_2^2$),
the matrix $Y^{\rm SM}= Y_1\, c_{\beta}+Y_2\, s_{\beta}$
is diagonalized by a biunitary transform $V_L^{u\dagger}Y^{\rm SM}V^u_R$ to $Y_D$,
with diagonal elements $y_i \equiv \sqrt2 m_i/\upsilon$.
However, the orthogonal combination
\begin{eqnarray}
\rho = V_L^{u\dagger}\left(-Y_1\, s_{\beta}+Y_2\, c_{\beta} \right)V^u_R,
\end{eqnarray}
cannot be simultaneously diagonalized,
and the exotic neutral Higgs bosons $H$ and $A$ possess 
FCNH couplings in general, including extra diagonal couplings $\rho_{ii}$,
\begin{equation}
-\sqrt{2}\mathcal{L}_Y
 = \bar{u}_{iL}\left[
   ({y_i\delta_{ij}}\, s_{\beta-\alpha} + {\rho_{ij}}\, c_{\beta-\alpha})\, h 
 + ({y_i\delta_{ij}}\, c_{\beta-\alpha} - {\rho_{ij}}\, s_{\beta-\alpha})\, H 
 -{i}\, \rho_{ij} \,\gamma_5\, A\right] u_{jR} + {\rm h.c.}
\end{equation}
The $\rho_{ij}$s are complex, i.e. $\arg\,\rho_{ij} \equiv \phi_{ij} \neq 0$,
and we have introduced the mixing angle $c_{\beta-\alpha}$ between
the two $CP$-even Higgs bosons $h$ and $H$.
It is known that the discovered 125 GeV boson $h$ is rather close to
the SM-Higgs boson, i.e. we are close to the \emph{alignment} limit~\cite{Gunion:2002zf} 
of $c_{\beta-\alpha} \to 0$ (hence $|s_{\beta-\alpha}| \to 1$). 
In this limit, the Yukawa couplings of $h$ are diagonal,
while $H$ and $A$ have FCNH couplings $\rho_{ij}$.
In the following, we largely adopt the alignment limit to simplify.

\section{EWBG}

Let us first give a heuristic account of EWBG.
Baryon number violation is facilitated by sphaleron processes in the symmetric phase.
As temperature cools, one has an expanding ``bubble'' of the broken phase.
But to avoid ``washout'' of the generated baryon number $n_B$
through the bubble wall, one needs $\Gamma_B^{\rm{(br)}}(T_C)<H(T_C)$,
i.e. the $n_B$ changing rate $\Gamma_B^{(\rm{br})}(T_C)$ is less than
the Hubble parameter $H(T_C)$ at critical temperature $T_C$.
This can be satisfied if the EWPT is first order such that
$\upsilon_C/T_C \gtrsim 1$, where $\upsilon_C^2 = \upsilon_1^2(T_C) + \upsilon_2^2(T_C)$.
This is where the ${\cal O}(1)$ Higgs couplings in 2HDM
differ from the rather weak Higgs self-coupling in SM,
that a strongly 1st order EWPT can be achieved through 
thermal loops involving extra Higgs bosons.

BAU, or $n_B/s \equiv Y_B \neq 0$, arises via
\begin{equation}
Y_B \equiv \frac{n_B}{s} =
 \frac{-3\Gamma_B^{(\rm{sym})}}{2D_q \lambda_+ s}
 \int_{-\infty}^0 dz' n_L(z') e^{-\lambda_-z'},
\end{equation}
where
 $\Gamma_B^{(\rm{sym})} = 120\alpha_W^5T$ is the $n_B$-changing rate in symmetric phase,
 $D_q \simeq 8.9/T$ is the quark diffusion constant, 
 $s$ is the entropy density,
 $\lambda_\pm \sim \upsilon_w$ is the bubble wall velocity, and
 $n_L$ is the total left-handed fermion number density.
The integration is over $z'$, the coordinate opposite bubble expansion direction.
We use the Planck value $Y_B^{\rm obs}= 8.59\times 10^{-11}$~\cite{Ade:2013zuv}
in our numerical analysis.

\begin{figure}[h]
\center
\includegraphics[width=3.2cm]{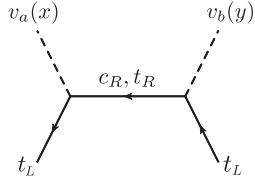}
\caption{
Leading CPV process for BAU, with bubble wall denoted 
symbolically as $\upsilon_{a}(x)$ and $\upsilon_b(y)$.
}
\label{fig:bubble_int}
\end{figure}

\vskip0.2cm
\noindent\underline{CPV Top Interactions}
\vskip0.1cm

Nonvanishing $n_L$ is needed for $Y_B$, which in our case is from the l.h. top density.
CPV interactions of (anti)top with the bubble wall 
is illustrated symbolically in Fig.~1,
where vertices can be read off from Eq.~(2.1).
The detailed ``transport'' problem is rather elaborate,
which we cannot possibly give full account here.
Suffice it to say that, 
with the closed time path formalism in the VEV insertion approximation,
the CPV source term $S_{ij}$ for left-handed fermion $f_{iL}$
induced by right-handed fermion $f_{jR}$ takes the form
\begin{equation}
{S_{i_L j_R}(Z)}= N_C F\,
    \rm{Im}\big[(Y_1)_{ij}(Y_2)_{ij}^*\big]\, v^2(Z)\, \partial_{t_Z}\beta(Z),
\end{equation}
where
 $Z = (t_Z,0,0,z)$ is position in heat bath (very early Universe),
 $N_C = 3$ is number of color, 
 $F$ is a function\footnote{
    See Ref.~\cite{Chiang:2016vgf} for explicit form, 
    as well as more details on the transport equations.}
  of complex energies of $f_{iL}$ and $f_{jR}$,
 and $\partial_{t_Z}\beta(Z)$ is the variation in $\beta(Z)$.
Note that, even though $\beta$ is basis-dependent in the general 2HDM,
its variation is physical and
plays an essential role in generating the CPV source term.
In our numerics, we take $\Delta\beta = 0.015$.

If bubble wall expansion and $\partial_{t_Z}\beta(Z)$
reflect departure from equilibrium, the essence of CPV for BAU is in the
$\rm{Im}\big[(Y_1)_{ij}(Y_2)_{ij}^*\big]$ factor in Eq.~(3.2).
Let us see how it depends on the couplings $\rho_{ij}$.
From Eq. (2.2) and the relation between $Y^{\rm SM}$ and $Y_D$, 
one has
\begin{equation}
\rm{Im}\big[(Y_1)_{ij}(Y_2)_{ij}^*\big]
= \rm{Im}\big[(V_L^uY_{\rm diag} V_R^{u\dagger})_{ij}(V_L^u\rho V_R^{u\dagger})_{ij}^* \big].
\end{equation}
To understand the result presented in the plot below,
suppose~\cite{Guo:2016ixx} $(Y_{1})_{tc} \neq 0$, $(Y_{2})_{tc} \neq 0$, 
and $(Y_1)_{tt}=(Y_2)_{tt} \neq 0$, while all else vanish
(we take $\tan\beta =1$ throughout for convenience).
Then $\sqrt{2}Y^{\rm SM} = Y_1 + Y_2$ can be 
diagonalized by just $V_R^u$ to a single nonvanishing 
$33$ element $y_t$, the SM Yukawa coupling,
while the combination $-Y_1 + Y_2$ is not diagonalized.
Solving for $V_R^u$ in terms of nonvanishing elements
in $Y_1$ and $Y_2$, one finds
\begin{equation}
  \rm{Im}\big[ (Y_1)_{tc}(Y_2)_{tc}^* \big]
 = -y_t\, \rm{Im}(\rho_{tt}), \quad \rho_{ct}=0,
\end{equation}
with $\rho_{tc}$ basically a free parameter.
Note that both doublets are involved for EWBG.

\begin{table*}[hb]
\center
{\small
\begin{tabular}{cccccc}
\hline\hline
$T_C \cong 119$ GeV & $\upsilon_C \cong 177$ GeV & $\upsilon_w \cong 0.4$
 & $\Delta\beta \cong 0.015$ & $D_q \cong 8.9/T$ & $D_H \cong 102/T$ \\
$m_{t_L}\cong0.59T$ & $m_{t_R}\cong0.62T$ & $m_{c_R}\cong0.50T$
 & $\Gamma_{q_{L,R}}\cong0.22T$ & $\Gamma_B^{(s)}\cong120\alpha_W^5T$ 
 & $\Gamma_{ss}=16\alpha_s^4 T$ \\
\hline\hline
\end{tabular}
}
\caption{
Input parameters for the $Y_B$ calculation.
}
\end{table*}

To calculate $n_L$ in Eq.~(3.1), one has a set of
transport equations that are diffusion equations fed 
by various density combinations weighted by mass (hence $T$) 
dependent statistical factors, as well as CPV source terms such as in Eq.~(3.2).
Following a relatively standard path, we reduce the coupled equations
to a single equation for $n_H$, controlled by a diffusion time 
$D_H \simeq 101.9/T$ modulated by $1/\upsilon_w^2$ 
(see Ref.~\cite{Chiang:2016vgf} for more discussion and references).
For experts, we list the important parameters in Table~1,
where one can see that $\upsilon_C/T_C \gtrsim 1$ is satisfied.

\begin{figure}[b]
\center 
\includegraphics[width=6cm]{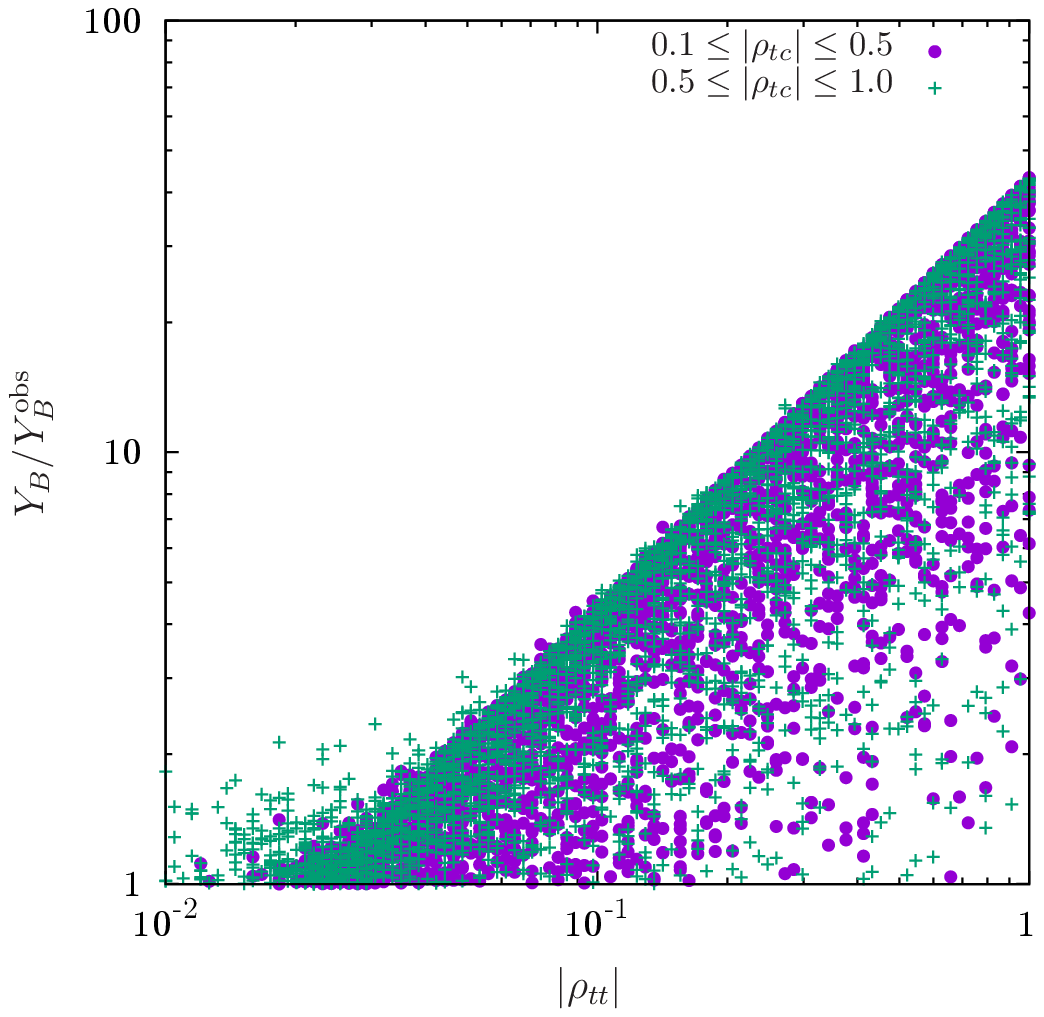} \hskip0.2cm
\includegraphics[width=6.15cm]{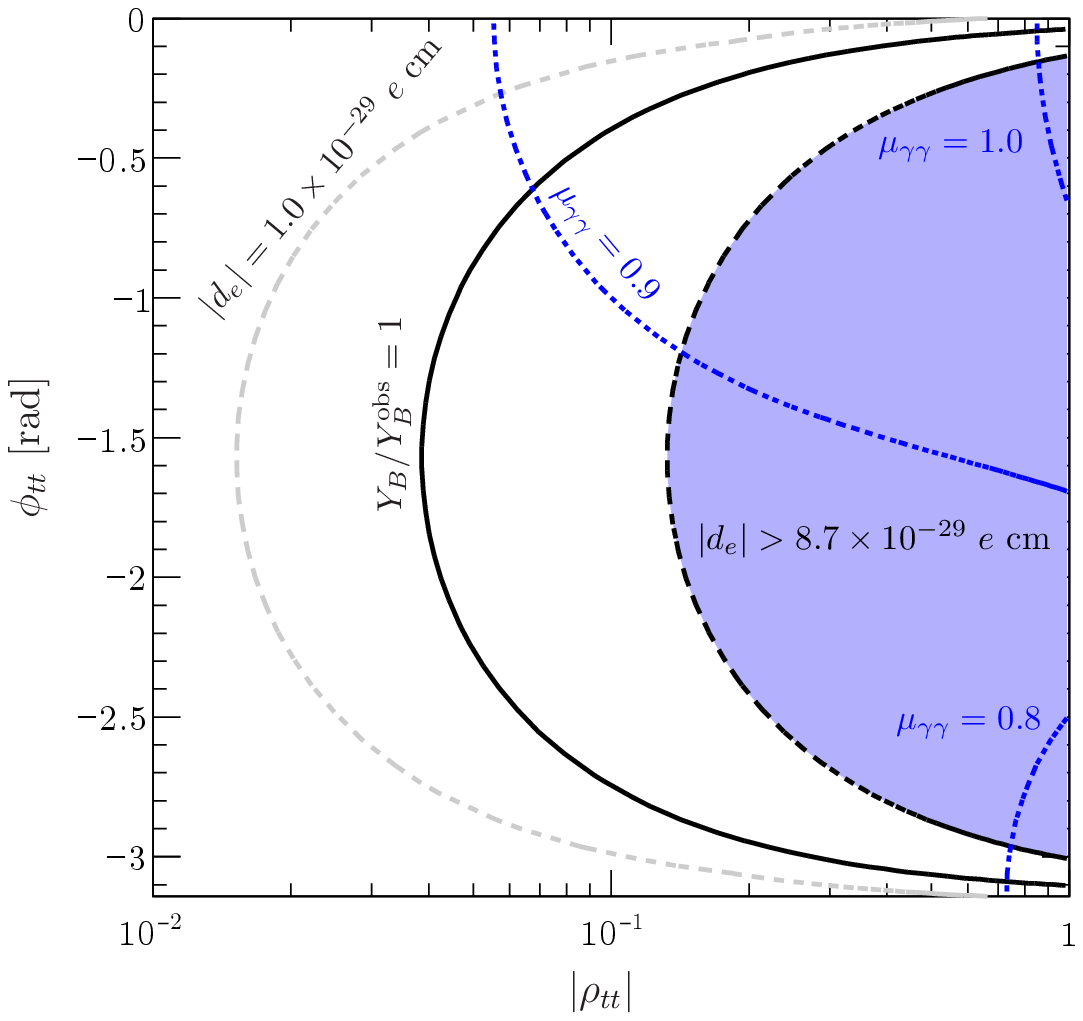}
\caption{
[left] $Y_B/Y_B^{\rm obs}$ vs $|\rho_{tt}|$, where purple dots (green crosses)
 are for $0.1\le |\rho_{tc}|\le 0.5$ ($0.5\le |\rho_{tc}|\le 1.0$);
[right] $Y_B/Y_B^{\rm{obs}}=1$ (solid) and $|d_e|$ (dashed)
 in the $|\rho_{tt}|$--$\phi_{tt}$ plane for $c_{\beta-\alpha} = 0.1$, 
 where shaded region is excluded,
 while dotted curves are for $h \to \gamma\gamma$ 
 with $\mu_{\gamma\gamma}$ as marked.
}
\end{figure}

Scanning over $|\rho_{tc}|$, $\phi_{tt}$ and $\phi_{tc}$,
we plot $Y_B/Y_B^{\rm obs}$ vs $|\rho_{tt}| \in (0.01,\, 1)$ in Fig.~2[left],
with $\rho_{tt}$ and $\rho_{tc}$ satisfying~\cite{Altunkaynak:2015twa} 
$B_d$, $B_s$ mixing as well as $b\to s\gamma$ constraints.
We have taken $m_H = m_A = m_{H^\pm} = 500$ GeV to simplify.
Though perhaps too restrictive, it illustrates the charm of EWBG:
the exotic Higgs masses are sub-TeV.
We separate $0.1\le|\rho_{tc}|\le 0.5$ and $0.5\le|\rho_{tc}|\le 1.0$,
which are plotted as purple dots and green crosses, respectively.
It is clear that sufficient $Y_B$ can be generated handsomely,
even for $|\rho_{tt}|$ considerably below 0.1.
Since no obvious difference is seen between lower and higher $|\rho_{tc}|$ 
for the bulk of the plot, we infer that $Y_B$ is driven by $\rho_{tt}$.
However, note that for small $|\rho_{tt}|$,
the green crosses populate $Y_B/Y_B^{\rm obs} \gtrsim 1$
much more than the purple dots, which means that $|\rho_{tc}| \sim {\cal O}(1)$
could take over EWBG for low $\rho_{tt}$, 
but it would demand near maximal $\phi_{tc}$. 

Thus, we have two mechanisms for BAU:
$\rho_{tt}$ as main driver, with $\rho_{tc}$ at ${\cal O}(1)$ as backup.

\section{Phenomenology}

As already mentioned, a leading effect is $t\to ch$ decay,
which demands $c_{\beta-\alpha} \neq 0$.
Taking $c_{\beta-\alpha} = 0.1$ and $|\rho_{tc}| = 1$,
we find ${\cal B}(t\to ch) = 0.15\%$, which satisfies
the latest ATLAS bound~\cite{Aaboud:2017mfd} of $0.22\%$
using 36.1 fb$^{-1}$ data at 13 TeV.
One recent motivation for FCNH was a 
hint for $h \to \tau\mu$ in 8 TeV data by CMS. 
Unfortunately, the hint disappeared with more data, 
and CMS sets a new bound~\cite{CMS:2017onh} of
${\cal B}(h \to \tau\mu) < 0.25\%$ based on 35.9 fb$^{-1}$ at 13 TeV.
Taking $c_{\beta-\alpha} = 0.1$, this still allows
${\cal B}(\tau \to \mu\gamma)$ up to $10^{-8}$,
which can be probed by Belle~II.

It is of interest to test CPV, as it links with EWBG.
Recent progress in the electron edm, $d_e$, by the ACME experiment
is rather astounding~\cite{Baron:2013eja}, 
which is shown ($c_{\beta-\alpha} = 0.1$) as 
the dashed curve in Fig.~2[right] with excluded region shaded, 
and $Y_B/Y_B^{\rm{obs}}=1$ given as the solid curve.
ACME projects an improvement by factor of 9 
(gray dashed curve), which could probe our EWBG mechanism.
The effect of $\rho_{tt}$ on $d_e$ given in Fig.~2[right], 
which is through the two-loop mechanism, assumes $\rho_{ee} = 0$.
For $|\rho_{ee}| \sim y_e = \sqrt2 m_e/\upsilon$ but purely imaginary,
cancellation between one- and two-loop effects could occur, 
allowing one to evade ACME.
What may be more exasperating is that the flavor or $CP$ violating effects 
mentioned so far would all vanish with $c_{\beta-\alpha} \to 0$, i.e. alignment.
What does not vanish with $c_{\beta-\alpha}$ is EWBG itself.
Nature seems skilled at producing the Universe, but 
hides the flavor and CPV traces.

EWBG in 2HDM needs both $\rho_{tt} \sim {\cal O}(1)$
and Higgs couplings $\sim {\cal O}(1)$.
We have also plotted in Fig.~2[right] possible reductions\footnote{
Loop effect from top via $\rho_{tt}$ could compensate~\cite{Hou:2017vvp} this reduction.
}
to $h \to \gamma\gamma$ width (dotted curves) due to $H^+$ effect,
which does not vanish with $c_{\beta-\alpha}$.
Another effect that does not vanish with alignment is
extra Higgs correction to $\lambda_{hhh}$, or triple-$h$ coupling,
which could receive 60\% enhancement with our benchmark,
$m_H = m_A = m_{H^\pm} = 500$ GeV.
The ``charm of EWBG", as mentioned, is of sub-TeV exotic scalars, 
which can be probed directly at LHC.
This is a consequence of ${\cal O}(1)$ self-couplings in the Higgs sector.
Of course, full degeneracy is clearly too restrictive,
and the actual parameter space should be much broader.
Together with the notion that $H^0$ and $A^0$ detection 
may be hampered by interference effects in $t\bar t$ decay final state,
search strategy for heavy Higgs should be readjusted.

\section{Conclusion}

We have studied~\cite{Fuyuto:2017ewj} EWBG induced by the top quark in 2HDM with FCNH couplings.
The leading effect arises from extra \emph{top} Yukawa coupling $\rho_{tt}$,
where BAU can be in the right ballpark for $\rho_{tt}\gtrsim0.01$
with moderate CPV phase $\phi_{tt}$.
Even if $|\rho_{tt}| \ll 0.1$, sufficient BAU can still be 
generated by $|\rho_{tc}|\simeq 1$ with large CPV phase $\phi_{tc}$.
These scenarios are testable in the future
with new flavor parameters that have rich implications,
and extra Higgs bosons below the TeV scale.

Nature may opt for a second Higgs doublet for generating the matter asymmetry of 
the Universe, through a new CPV phase associated with the top quark.
As a bonus, it is found that alignment emerges naturally 
from such 2HDM without discrete symmetry~\cite{Hou:2017hiw}.

\end{document}